\documentclass{Rinton-P9x6}

\def\lsim{\mathrel{\raise.3ex\hbox{$<$\kern-.75em\lower1ex\hbox{$\sim$}}}}
\def\gsim{\mathrel{\raise.3ex\hbox{$>$\kern-.75em\lower1ex\hbox{$\sim$}}}}
\newcommand\openone{\leavevmode\hbox{\small1\normalsize\kern-.33em1}}

\begin{document}

\title{Phenomenology of the Littlest Higgs Model}

\author{Heather E. Logan}

\address{Department of Physics,
University of Wisconsin, 1150 University Avenue, Madison, Wisconsin 53706\\
E-mail: logan@pheno.physics.wisc.edu
}

%


\maketitle

\abstracts{
The little Higgs idea is a new way to solve the little hierarchy
problem by protecting the Higgs mass from quadratically divergent
one-loop corrections.  In this talk I describe the phenomenology 
of one particular realization of the little Higgs idea: the 
``Littlest Higgs'' model.}

\section{Introduction}

One of the major motivations for physics beyond the Standard Model (SM)
is to resolve the hierarchy and fine-tuning problems between the 
electroweak scale and the Planck scale.
The Higgs boson mass in the SM is quadratically sensitive
to the cutoff scale $\Lambda$ of the SM effective theory
via radiative corrections.
The quantum-corrected Higgs mass is given at one-loop by
\begin{equation}
	m_h^2 = \left( m_h^2 \right)_{\rm bare} 
	+ \frac{3 g^2 \Lambda^2}{32 \pi^2 m_W^2}
	\left( m_h^2 + 2 m_W^2 + m_Z^2 - \frac{4}{3} N_c \sum_f m_f^2
	\right).
	\label{eq:mhcorr}
\end{equation}
For a high cutoff scale $\Lambda$, this cancellation must be
fine-tuned; for example, for $\Lambda = 10$ TeV, $(m_h^2)_{\rm bare}$ 
must be tuned at the 1\% level to cancel the radiative corrections.
In fact, requiring that
the one-loop contributions to the Higgs mass-squared parameter are no
more than 10 times the size of the renormalized Higgs mass-squared term
(i.e., no more than 10\% fine-tuning), leads to the requirement that
\begin{equation}
	\Lambda_t \lsim 2 \ {\rm TeV}, \qquad
	\Lambda_{W,Z} \lsim 5 \ {\rm TeV}, \qquad
	\Lambda_H \lsim 10 \ {\rm TeV}.
\end{equation}

What could the cancellation mechanism be?  The classic solution is
supersymmetry.
From the bottom-up point of view, the quadratic 
divergences in the Higgs mass due to top quark, gauge boson and Higgs
loops are canceled by the top squark, gaugino and Higgsino loops,
respectively.  From the top-down point of view, the Higgs
mass is protected by supersymmetry to be one loop factor below the soft
supersymmetry breaking scale.
Thus weak scale supersymmetry is natural if 
$M_{SUSY} \sim \mathcal{O}(1 \ {\rm TeV})$.

The little Higgs idea \cite{bigmoose}
is an alternative way to keep the Higgs boson
naturally light.  The basic idea is as follows (see Fig.~\ref{fig:scales}):  
\begin{figure}
\begin{center}
\resizebox{10cm}{!}{\includegraphics{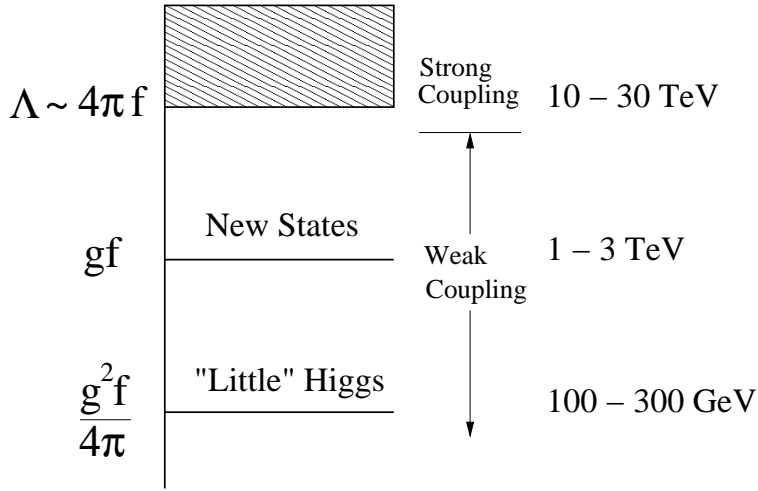}}
\end{center}
\caption{An illustration of the scales in the little Higgs picture.
From $^2$.} 
\label{fig:scales}
\end{figure}

(i) The Higgs field is
a pseudo-Nambu-Goldstone boson \cite{PNGBhiggs} of a global symmetry
that is spontaneously broken at a scale $\Lambda \sim 4 \pi f \sim 10-30$ TeV;

(ii) The quadratic divergences in the Higgs mass are canceled at the
one-loop level by new particles with masses $M \sim g f \sim 1-3$ TeV;

(iii) The Higgs acquires a mass radiatively at the electroweak scale
$v \sim g^2 f/4 \pi \sim 100-300$ GeV.

From the bottom-up point of view, the quadratic divergences in the
Higgs mass are canceled by loops of new particles of the same statistics
(in contrast to supersymmetry, in which the cancellations are due to particles
of opposite statistics).
From the top-down point of view, the Higgs mass is protected by the
global symmetry.  Little Higgs models 
\cite{bigmoose,minmoose,Littlest,SU6Sp6,KaplanSchmaltz,custodialmoose,SkibaTerning,Spencer}
are constructed so that at least two
operators are needed to explicitly break all of the global symmetry
that protects the Higgs mass.  This forbids quadratic divergences at
one-loop; the Higgs mass is then smaller than $\Lambda$ by not one 
but two loop factors, leading to the little hierarchy 
$\Lambda \gg f \gg v$.

In this talk I start by reviewing the Littlest Higgs model \cite{Littlest}
in Sec.~\ref{sec:model}, then give an overview of its phenomenology
in Sec.~\ref{sec:pheno}.  Section~\ref{sec:concl} contains an outlook
and conclusions.  This talk is based on \cite{LHpheno,LHloop}.

\section{\label{sec:model} The Littlest Higgs Model}

To study phenomenology we must specify a model.  We consider here a
specific realization of the little Higgs idea called the ``Littlest
Higgs'' model \cite{Littlest}, introduced last year by Arkani-Hamed, 
Cohen, Katz and Nelson.

The Littlest Higgs model is a nonlinear sigma model
with a global SU(5) symmetry group broken down to SO(5) by the 
vacuum expectation value (vev) 
\begin{equation}
	\Sigma_0 = \left(
          \begin{array}{ccc}
            \quad & \quad & \openone \\ \quad & 1 &\quad \\ \openone
            &\quad&\quad
          \end{array}
          \right).
\end{equation}
An [SU(2)$\times$U(1)]$^2$ subgroup of SU(5) is gauged; $\Sigma_0$
breaks this gauge symmetry down to the diagonal SU(2)$\times$U(1) subgroup,
which is identified with the SM gauge group.  The symmetry breaking
leads to $24-10 = 14$ Goldstone bosons, four of which are eaten by
the broken gauge generators.  The remaining ten Goldstone bosons 
transform under the
SM gauge symmetry as a complex doublet $h$ (which will become the SM
Higgs doublet) and a complex triplet $\phi$.  The Goldstone bosons can
be written as $\Sigma = \exp(2 i \Pi / f) \Sigma_0$, where 
$f \sim \Lambda/4\pi$ is the ``pion decay constant'' that will
set the scale of the new particle masses. The uneaten Goldstone
bosons are given by
\begin{equation}
            \Pi = \left(
            \begin{array}{ccc}
                \quad & h^\dagger/\sqrt{2} & \phi^\dagger \\
                h/\sqrt{2} & \quad & h^\star/\sqrt{2} \\
                \phi & h^{\rm T}/\sqrt{2} & \quad
            \end{array}
            \right),\quad
           h = (h^+,\ h^0),\quad
            \phi = \left(
            \begin{array}{cc}
                \phi^{++} & \phi^+/\sqrt{2}  \\
                \phi^+/\sqrt{2} & \phi^0
            \end{array}
            \right).
\end{equation}

\subsection{Gauge sector}

The gauge couplings break the global symmetry explicitly.
However, the model is constructed such that no single interaction 
breaks all the global symmetry protecting the Higgs mass.  This 
implements the little Higgs mechanism: at least two interactions are
required to break all the global symmetry and give mass to the Higgs,
thus forbidding quadratically divergent radiative corrections at the
one loop level.

The gauge generators are chosen to be
\begin{eqnarray}
	Q_1^a = \left( \begin{array}{ccc}
                \sigma^a/2 & \quad & \quad \\
                \quad & \quad & \quad \\
                \quad & \quad & \quad
                \end{array} \right),
	& \qquad &
	Q_2^a = \left( \begin{array}{ccc}
                \quad & \quad & \quad \\
                \quad & \quad & \quad \\
                \quad & \quad & -\sigma^a/2
                \end{array} \right),
	\nonumber \\
	Y_1 = {\rm diag}(-3, -3, 2, 2, 2)/10, & \qquad &
	Y_2 = {\rm diag}(-2, -2, -2, 3, 3)/10,
\end{eqnarray}
with gauge couplings $g_1,g_2,g_1^{\prime}$, and $g_2^{\prime}$,
respectively.
The generators $Q_1^a$ and $Y_1$ preserve a global SU(3)$_1$ symmetry
while the generators $Q_2^a$ and $Y_2$ preserve a second global
SU(3)$_2$ symmetry, each of which forbids a mass for $h$:
\begin{equation}
	SU(3)_1 = \left( \begin{array}{c|ccc}
            0_{2\times2} & & & \\
                         \hline
                         & & & \\
                         & & V_3 & \\
                         & & & \\
            \end{array}\right), \qquad \qquad
        SU(3)_2 = \left( \begin{array}{ccc|c}
                         & & & \\
                         & V_3 & & \\
                         & & & \\
                         \hline
                         & & & 0_{2\times2}\\
            \end{array}\right).
	\label{eq:SU3global}
\end{equation}

The $\Sigma_0$ vev gives mass to one linear combination of the two SU(2)
gauge bosons $W_1$ and $W_2$ and to one linear combination of the two
U(1) gauge bosons $B_1$ and $B_2$ as follows:
\begin{eqnarray}
	& W^{\prime a} = - c W_1^a + s W_2^a, \qquad \qquad 
		& M_{W^{\prime}} = \frac{g f}{2 s c},
	\nonumber \\
	& B^{\prime} = - c^{\prime} B_1 + s^{\prime} B_2, \qquad \qquad
		& M_{B^{\prime}} = \frac{g^{\prime} f}
		{2\sqrt{5} s^{\prime} c^{\prime}},
\end{eqnarray}
where we define the mixing angles $c,s \equiv \cos \theta, \sin\theta$
and $c^{\prime},s^{\prime} \equiv \cos\theta^{\prime}, \sin\theta^{\prime}$
in terms of the gauge couplings by
\begin{equation}
	g = g_1 s = g_2 c, \qquad \qquad 
	g^{\prime} = g_1^{\prime} s^{\prime} = g_2^{\prime} c^{\prime}.
\end{equation}

The cancellation of the quadratic divergence in the Higgs mass from 
gauge boson loops can now be seen explicitly by examining the Lagrangian.
The couplings of gauge boson pairs to $h^{\dagger}h$ arises only through
collective breaking, which ensures the cancellation of the divergence:
\begin{eqnarray}
	\mathcal{L} &=& \frac{1}{4} \left( g_1 g_2 W_1^a W_2^a
	+ g_1^{\prime} g_2^{\prime} B_1 B_2 \right) h^{\dagger} h
	+ \cdots
	\nonumber \\
	&=& \frac{1}{4} \left[ g^2 \left( W^a W^a - W^{\prime a} W^{\prime a}
	\right) + g^{\prime 2} \left( B B - B^{\prime} B^{\prime} \right)
	\right] h^{\dagger} h + \cdots.
\end{eqnarray}
The coupling of $h^{\dagger}h$ to pairs of heavy and light gauge bosons 
are equal in magnitude and opposite in sign,
leading to the cancellation of the quadratic divergence from the SM 
gauge boson loops by the corresponding heavy gauge bosons.

After electroweak symmetry breaking, the heavy gauge bosons mix with
the SM gauge bosons at order $v^2/f^2$ and form the mass eigenstates
$W_H,Z_H,A_H$.

\subsection{Top sector}

To cancel the quadratic divergence due to the top quark loop, the Higgs
coupling to the top quark must also be generated through collective breaking.
This can be done by introducing a vector-like pair of colored Weyl 
fermions $\tilde t$ and $\tilde t^{\prime c}$, and writing the 
following couplings:
\begin{equation}
	\mathcal{L}_{\rm Yuk} = \frac{\lambda_1}{2} f \epsilon_{ijk}
	\epsilon_{xy} Q_i \Sigma_{jx} \Sigma_{ky} u_3^{\prime c}
	+ \lambda_2 f \tilde t \tilde t^{\prime c} + {\rm h.c.},
	\label{eq:Lyuk}
\end{equation}
where the third-generation quark doublet is expanded to 
$Q = (b, t, \tilde t)$ \cite{Littlest}.  (For an alternative top 
sector, see \cite{Anntalk}.)
Here $i,j,k = 1,2,3$ and $x,y = 4,5$.
The first term in Eq.~\ref{eq:Lyuk} generates the Higgs couplings
to fermions when $\Sigma$ is expanded in powers of the Goldstone bosons.
It is symmetric under the 
global SU(3)$_2$ symmetry defined in Eq.~\ref{eq:SU3global}, thus 
ensuring that the quadratic divergences cancel between the top loop
and a loop of the new heavy fermion.  The second term in Eq.~\ref{eq:Lyuk}
is a mass term for the vector-like quark.  This term preserves
the global SU(3)$_1$ symmetry of Eq.~\ref{eq:SU3global}.  Inserting the
$\Sigma_0$ vev, $\tilde t$ marries a linear combination of 
$\tilde t^{\prime c}$ and $u_3^{\prime c}$
and gets a mass of order $f$,
\begin{equation}
	M_T = f \sqrt{\lambda_1^2 + \lambda_2^2}.
\end{equation}
The remaining linear combination becomes the right-handed top quark.
The top quark mass is given by
\begin{equation}
	m_t = \frac{\lambda_1 \lambda_2}{\sqrt{\lambda_1^2 + \lambda_2^2}} v.
\end{equation}
Note that $m_t$ vanishes if either $\lambda_1$ or $\lambda_2$ is zero:
this is a manifestation of the collective breaking.

\subsection{\label{sec:fermions} The rest of the fermions}

There is no need to cancel the quadratic divergences in the 
Higgs mass due to light fermion loops because they 
do not become important until scales much
higher than the 10 TeV cutoff of the nonlinear sigma model.
Thus we can generate masses for the rest of the fermions by writing 
terms of the same form as Eq.~\ref{eq:Lyuk}, but without the extra vector-like
quarks.  For the down-type fermion masses, $\Sigma$ is replaced by $\Sigma^*$.

If we make this choice for the light fermion masses,
then gauge invariance of the Lagrangian in
Eq.~\ref{eq:Lyuk} fixes the charges of the fermions under the
two U(1) gauge symmetries up to only two free continuous parameters
$y_u$ and $y_e$ per generation (see Table~\ref{tab:U1charges}).
\begin{table}
\begin{tabular}{|c||c|c|c|c|c||c|c|}
\hline
 & $Q$ & $u^{\prime c}$ & $d^c$ & $L$ & $e^c$ 
	& $\tilde t$ & $\tilde t^{\prime c}$ \\
\hline
$Y_1$ & $-\frac{3}{10} - y_u$ & $y_u$ & $\frac{3}{5} + y_u$
                        & $\frac{3}{10} - y_e$ & $y_e$
                        & $\frac{1}{5} - y_u$ & $-\frac{1}{5} + y_u$ \\
                \hline
                $Y_2$ & $\frac{7}{15} + y_u$ & $-\frac{2}{3} - y_u$
                        & $-\frac{4}{15} - y_u$
                        & $-\frac{4}{5} + y_e$ & $1-y_e$
                        & $\frac{7}{15} + y_u$ & $-\frac{7}{15} - y_u$ \\
                \hline
            \end{tabular}
\caption{Fermion charges under the two U(1) gauge symmetries.}
\label{tab:U1charges}
\end{table}
Imposing anomaly cancellation then fixes the U(1) charges uniquely,
requiring:
\begin{equation}
	y_u = -2/5, \qquad \qquad y_e = 3/5.
\end{equation}
Different U(1) charges are possible for the light fermions if their masses
are instead generated by higher-dimensional operators.

\subsection{Higgs potential and electroweak symmetry breaking}

The Higgs potential is generated radiatively by integrating out the 
heavy gauge bosons and heavy top-partner.  
It can be written in the general form,
\begin{equation}
	V = \lambda_{\phi^2} f^2 {\rm Tr}(\phi^{\dagger} \phi)
                + i \lambda_{h \phi h} f \left( h \phi^{\dagger} h^T
                        - h^* \phi h^{\dagger} \right)
                - \mu^2 h h^{\dagger}
                + \lambda_{h^4} ( h h^{\dagger} )^2.
\end{equation}
The one-loop contributions to $\lambda_{\phi^2}$, $\lambda_{h\phi h}$
and $\lambda_{h^4}$ are quadratically sensitive to the cutoff $\Lambda$.
We thus introduce order-one coefficients $a,a^{\prime}$ to parameterize our
ignorance of cutoff-scale physics in the gauge and fermion loops, respectively.
We then have,
\begin{eqnarray}
        \lambda_{\phi^2} &=& \frac{a}{2} \left[ \frac{g^2}{s^2c^2}
        + \frac{g^{\prime 2}}{s^{\prime 2}c^{\prime 2}} \right]
        + 8 a^{\prime} \lambda_1^2, \nonumber \\
        \lambda_{h \phi h} &=& -\frac{a}{4}
        \left[ g^2 \frac{(c^2-s^2)}{s^2c^2}
        + g^{\prime 2} \frac{(c^{\prime 2}-s^{\prime 2})}
        {s^{\prime 2}c^{\prime 2}} \right]
        + 4 a^{\prime} \lambda_1^2, \nonumber \\
        \lambda_{h^4} &=& \frac{a}{8} \left[ \frac{g^2}{s^2c^2}
        + \frac{g^{\prime 2}}{s^{\prime 2}c^{\prime 2}} \right]
        + 2 a^{\prime} \lambda_1^2 = \frac{1}{4} \lambda_{\phi^2}.
\end{eqnarray}
The parameter $\mu^2$ gets log-divergent contributions at one-loop
and quadratically divergent contributions at two-loop, which are 
both generically one loop factor smaller than $f^2$, leading to a
Higgs mass of order $g^2 f/4 \pi$.  Since there will be 
additional cutoff-scale freedom in the two-loop contributions to 
$\mu^2$, we regard it as a free parameter.  For $\mu^2 > 0$, electroweak
symmetry is broken, and both $h$ and $\phi$ get vevs:
\begin{equation}
	2 \langle h^0 \rangle^2 \equiv v^2 = \frac{\mu^2}{\lambda_{h^4} - 
	\lambda^2_{h\phi h} / \lambda_{\phi^2}}, \qquad \qquad
	\langle i \phi^0 \rangle \equiv v^{\prime} = 
	\frac{\lambda_{h\phi h}}{2 \lambda_{\phi^2}} \frac{v^2}{f}.
\end{equation}
Note that $v^{\prime} \sim v^2/f$ is much smaller than $v$.

The scalar masses, to leading order in $v/f$, are
\begin{equation}
	M^2_{\phi} = \lambda_{\phi^2} f^2, \qquad \qquad
	m^2_H = 2 \mu^2.
\end{equation}

\subsection{Summary of new parameters and particles}

The Littlest Higgs model contains six new free parameters, which we 
can choose as follows:

(1) $\tan \theta = s/c = g_1/g_2$: new SU(2) gauge coupling.

(2) $\tan \theta^{\prime} = s^{\prime}/c^{\prime} 
	= g_1^{\prime}/g_2^{\prime}$: new U(1) gauge coupling.

(3) $f$: symmetry breaking scale, $\mathcal{O}$(TeV).

(4) $v^{\prime}$: triplet vev; $v^{\prime} < v^2/4f$.

(5) $m_H$: SM Higgs mass.

(6) $M_T$: top-partner mass (together with $m_t$, this fixes
	$\lambda_1$ and $\lambda_2$).

The new particles and their masses are summarized in Table~\ref{tab:particles}.
\begin{table}
\begin{center}
\begin{tabular}{|c|cccc|}
\hline
Particle & $A_H$ & $Z_H,W_H$ & $\Phi^0,\Phi^P,\Phi^+,\Phi^{++}$ & $T$ \\
\hline
Mass & $f \frac{m_Z s_W}{v\sqrt{5}s^{\prime}c^{\prime}}$ \qquad
	& $f \frac{m_W}{vsc}$ \qquad
	& $f \frac{\sqrt{2} m_H}{v \sqrt{1 - (4v^{\prime}f/v^2)^2}}$ \qquad
	& $f \sqrt{\lambda_1^2 + \lambda_2^2}$ \\
\hline
Mass lower bound & $0.16 f$ & $0.65 f$ & $0.66 f$ & $1.42 f$ \\
\hline
\end{tabular}
\end{center}
\caption{The new particles of the Littlest Higgs model and their masses
to leading order in $v/f$.  The masses given all receive corrections
of order $v^2/f$.  
For $M_{\phi}$, we obtain the lower bound by assuming $m_H \geq 115$ GeV.}
\label{tab:particles}
\end{table}

\section{\label{sec:pheno} Phenomenology}

There are by now quite a number of little Higgs models in the 
literature 
\cite{bigmoose,minmoose,Littlest,SU6Sp6,KaplanSchmaltz,custodialmoose,SkibaTerning,Spencer}.  
It thus behooves us to look for generic features of the phenomenology.
All little Higgs models must contain the following features at the TeV scale:
\begin{itemize}
	\item New heavy gauge bosons to cancel the $W$ and $Z$ loops.
	In the Littlest Higgs model these are almost pure SU(2) gauge
	bosons.  The SM fermions must transform under only one
	of the two gauged SU(2) symmetries -- say, SU(2)$_1$ -- so their
	couplings to the new heavy gauge bosons are universal, 
	$\propto g \cot\theta$.  This determines the cross section
	for Drell-Yan production of the heavy gauge bosons at the LHC.
	This appears to be a generic feature of 
	``product gauge group'' little Higgs models 
	\cite{bigmoose,minmoose,Littlest,SU6Sp6,custodialmoose,Spencer}
	in which the SU(2)$_L$ SM gauge group comes from
	the breaking of two groups down to a diagonal subgroup.
	The decays of the heavy gauge bosons are more
	model dependent, since non-fermionic decays can play a role.
	\item A new heavy fermion to cancel the top quark loop.
	The production and decay modes of the heavy fermion are fixed by the
	form of the Yukawa Lagrangian, Eq.~\ref{eq:Lyuk}, which appears
	in many little Higgs models.  
	Some models contain more than one top-partner and/or partners
	for the two light fermion generations, although typically only
	one of these states cancels the top loop divergence.
	\item New heavy scalars to cancel the Higgs loop.
	The heavy scalar sector is very model dependent, and can
	consist of singlets, doublets or triplets.  Some models have
	two light Higgs doublets 
	\cite{minmoose,SU6Sp6,KaplanSchmaltz,custodialmoose,SkibaTerning}.
\end{itemize}

\subsection{Electroweak precision constraints}

The constraints on little Higgs models from electroweak precision 
data have been examined in detail in 
\cite{GrahamEW1,JoAnneEW,GrahamEW2,JayEW}.  
The constraints come from $Z$ pole data, low-energy neutrino-nucleon 
scattering, and the $W$ mass measurement.  Together, these measurements
probe little Higgs model contributions from the exchange of the 
heavy gauge bosons between fermion pairs, mixing between the heavy
and light gauge bosons that modifies the Z boson couplings to fermions,
and a shift in the mass ratio of the $W$ and $Z$.

The contributions to the various observables in the Littlest Higgs
model are outlined in
Table~\ref{tab:EWpulls}.
\begin{table}
\begin{center}
\begin{tabular}{|c|c|c|c|}
\hline
  & $SU(2)_H$ & $U(1)_H$ & $\langle \phi \rangle$ \\
\hline
$M^2_W$ & $-\frac{5}{12} \frac{v^2}{f^2} + c^2s^2 \frac{v^2}{f^2}$
&
$0$ & $4
\frac{v^{\prime 2}}{v^2}$ \\
\hline
$M^2_Z$ & $-\frac{5}{12} \frac{v^2}{f^2} + c^2s^2 \frac{v^2}{f^2} $ &
$-\frac{5}{4}
\frac{v^2}{f^2} (c^{\prime2} -s^{\prime2})^2$ & $8 \frac{v^{\prime
2 }}{v^2}$ \\
\hline
$G_F$ &  $\frac{5}{12} \frac{v^2}{f^2}$ & $0$ & $ -4
\frac{v^{\prime 2}}{v^2}$ \\
\hline
$M^2_Z G_F$ &  $c^2 s^2 \frac{v^2}{f^2} $ & $-\frac{5}{4}
\frac{v^2}{f^2}(c^{\prime 2}-s^{\prime2})^2$ & $4
\frac{v^{\prime 2}}{v^2}$ \\
\hline
$\delta{g_{ff}}$ &  $\propto c^{2} \frac{v^2}{f^2} $ & $
\propto (-c^{\prime 2}Y_1+s^{\prime2}Y_2) \frac{v^2}{f^2} $ & $0$   \\
\hline
\end{tabular}
\end{center}
\caption{Extra contributions to the electroweak parameters in the 
Littlest Higgs model from the exchange of heavy SU(2) and U(1) gauge 
bosons and from the triplet vev.  $\delta g_{ff}$ collectively
denotes the modification of the neutral current couplings of the SM 
fermions.}
\label{tab:EWpulls}
\end{table}
Examining these contributions yields a strategy for reducing the impact of
the Littlest Higgs model on electroweak observables:

	(1) Reduce the triplet vev: $v^{\prime} \ll v$;

	(2) Reduce the heavy SU(2) gauge boson contributions: $c \ll 1$;

	(3) Reduce the heavy U(1) gauge boson contributions to $M_Z$:
	$c^{\prime} \approx s^{\prime}$;

	(4) Reduce the heavy U(1) gauge boson contributions to neutral
	current couplings:
	$c^{\prime 2} Y_1 \approx s^{\prime 2} Y_2$.

The first three conditions can be straightforwardly satisfied by going
to the appropriate parameter region.  The fourth condition is more 
difficult, since it depends on the U(1) charge assignments of (mainly the
first two generations of) fermions.  It turns out that the charge 
assignments obtained from requiring that the light fermion mass terms
have the same form as those of the third generation (Eq.~\ref{eq:Lyuk})
and imposing anomaly
cancellation satisfy the fourth condition reasonably well.  These 
lead to a lower bound on $f$ of about 1 TeV \cite{GrahamEW2},
corresponding to a lower bound on $M_{W_H},M_{Z_H}$ of about 2 TeV.
The cancellation
in the fourth condition can be improved by requiring the light fermion
masses to be generated by appropriate 
higher-dimensional operators so that their 
U(1) charges can be chosen with more freedom; however, this is not 
really necessary to avoid fine-tuning.

The situation can also be simplified
by gauging only one U(1) symmetry (i.e., hypercharge), so that 
there is no heavy U(1) gauge boson and the second column in 
Table~\ref{tab:EWpulls} is eliminated.
Removing the heavy U(1) gauge boson has the added benefit of avoiding
constraints \cite{JoAnneEW} from direct Tevatron $Z^{\prime}$ searches.

These conclusions can be generalized to other little Higgs models.
The lower bounds on the masses of the new heavy gauge bosons are 
generally in the $1.5-2$ TeV range \cite{GrahamEW2,JayEW}.
In models with a product gauge group, the electroweak precision measurements
favor parameter regions in which the new
heavy gauge bosons are approximately decoupled from the SM fermions;
e.g., $\cot\theta \simeq 0.2$.

The electroweak precision measurements do not directly constrain the mass 
of the top-partner.  However, the mass of the top-partner is related to 
the heavy gauge boson masses by the structure of the model.  For naturalness, 
the top-partner should be as light as possible.  The lower bounds on the
top-partner mass are generally in the $1-2$ TeV range.

\subsection{Collider signatures}

\begin{figure}
\begin{center}
\resizebox{10cm}{!}{\includegraphics{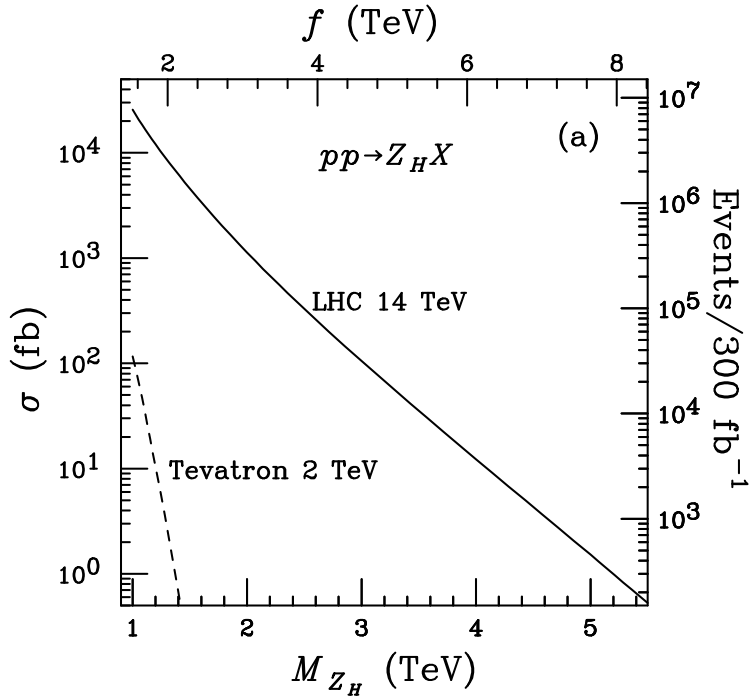}}
\end{center}
\caption{Cross section for $Z_H$ production in Drell-Yan at the LHC
and Tevatron, for $\cot\theta = 1$.  From $^{11}$.}
\label{fig:sigmaZH}
\end{figure}

{\it $Z_H$ and $W_H$:}
The heavy SU(2) gauge bosons $Z_H$ and $W_H$ can be produced via Drell-Yan
at the LHC (and at the Tevatron, if they are light enough).
The cross section is proportional to $\cot^2\theta$ because the 
$Z_H$ and $W_H$ couplings to fermion pairs are proportional to $\cot\theta$.
In Fig.~\ref{fig:sigmaZH} we show the cross section for $Z_H$ production
at the Tevatron and LHC for $\cot\theta = 1$.
In the region of small $\cot\theta \simeq 0.2$ favored by the precision 
electroweak data, the cross section must be scaled down by 
$\cot^2\theta \simeq 0.04$.  Even with this suppression factor, 
a cross section of 40 fb is expected at the LHC for $M_{Z_H} \simeq 2$ TeV,
leading to 4,000 events in 100 fb$^{-1}$ of data.
The production and decay of $Z_H$ and $W_H$ at the LHC has also been studied
in \cite{gustavo}.

\begin{figure}
\begin{center}
\resizebox{10cm}{!}{\includegraphics{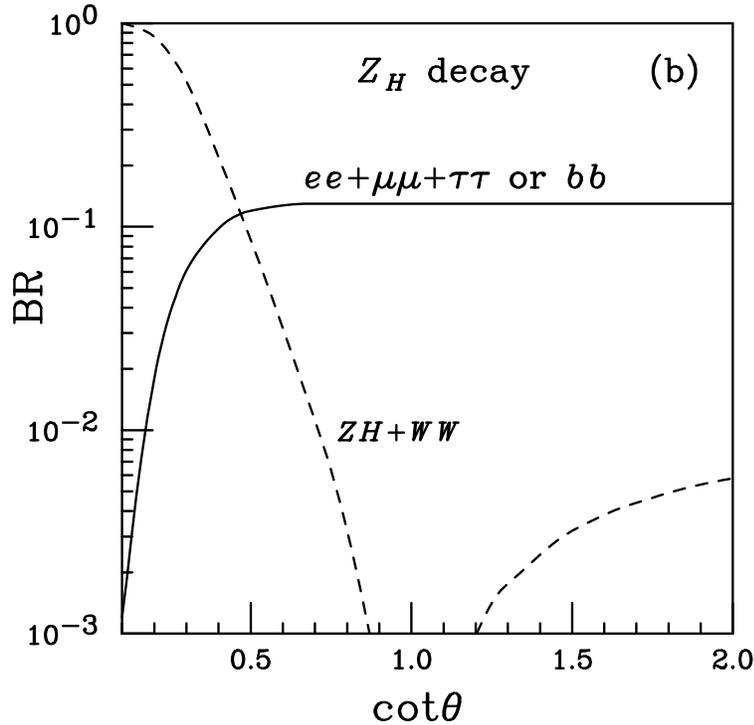}}
\end{center}
\caption{Branching ratios of $Z_H$ into SM particles as a function
of $\cot\theta$, neglecting final-state mass effects.}
\label{fig:brZH}
\end{figure}

The decay branching fractions of $Z_H$ are shown in 
Fig.~\ref{fig:brZH}.\footnote{Here we correct an error in 
\cite{LHpheno,gustavo} in which the $Z_H \to W^+W^-$ decay mode was
overlooked.}
The partial widths to fermion pairs are proportional to $\cot^2\theta$,
while the partial widths to $ZH$ and $W^+W^-$ are proportional to 
$\cot^2 2\theta$.  This offers a method to distinguish the Littlest
Higgs model from a ``big Higgs'' model with the same gauge group in which
the Higgs doublet transforms under only one of the SU(2) groups 
\cite{gustavo}, in which case the $ZH$ and $W^+W^-$ partial widths
would also be proportional to $\cot^2\theta$.
Neglecting final-state particle masses, the branching fraction into three
flavors of charged leptons is equal to that into one flavor of quark
($\simeq 1/8$ for $\cot\theta \gsim 0.5$), due 
to the equal coupling of $Z_H$ to all SU(2) fermion doublets.  The branching
ratio into $ZH$ is equal to that into $W^+W^-$.
The total width of $Z_H$ depends on $\cot\theta$; for $\cot\theta \sim 0.2$
the $Z_H$ width is about 1\% of the $Z_H$ mass.

The $W_H^{\pm}$ couplings to fermion doublets are larger by a factor of
$\sqrt{2}$ than the $Z_H$ couplings; this together with the parton distribution
of the proton leads to a $W_H^{\pm}$ cross
section at the LHC about 1.5 times that of $Z_H$ \cite{gustavo}.
As for the $W_H$ decays, the branching fraction into three lepton flavors
is equal to that into one generation of quarks ($\simeq 1/4$ for $\cot\theta
\gsim 0.5$).  At low $\cot\theta$, $W_H^{\pm}$ decays predominantly into
$W^{\pm} H$ and $W^{\pm} Z$ with partial widths proportional to 
$\cot^2 2\theta$.

The general features of the production and decay of $Z_H$ and $W_H$
should extend to other little Higgs models with ``product gauge groups''.
The decay to $ZH$ will be modified in models that 
contain more than one light Higgs doublet.

\begin{figure}
\begin{center}
\resizebox{10cm}{!}{\includegraphics{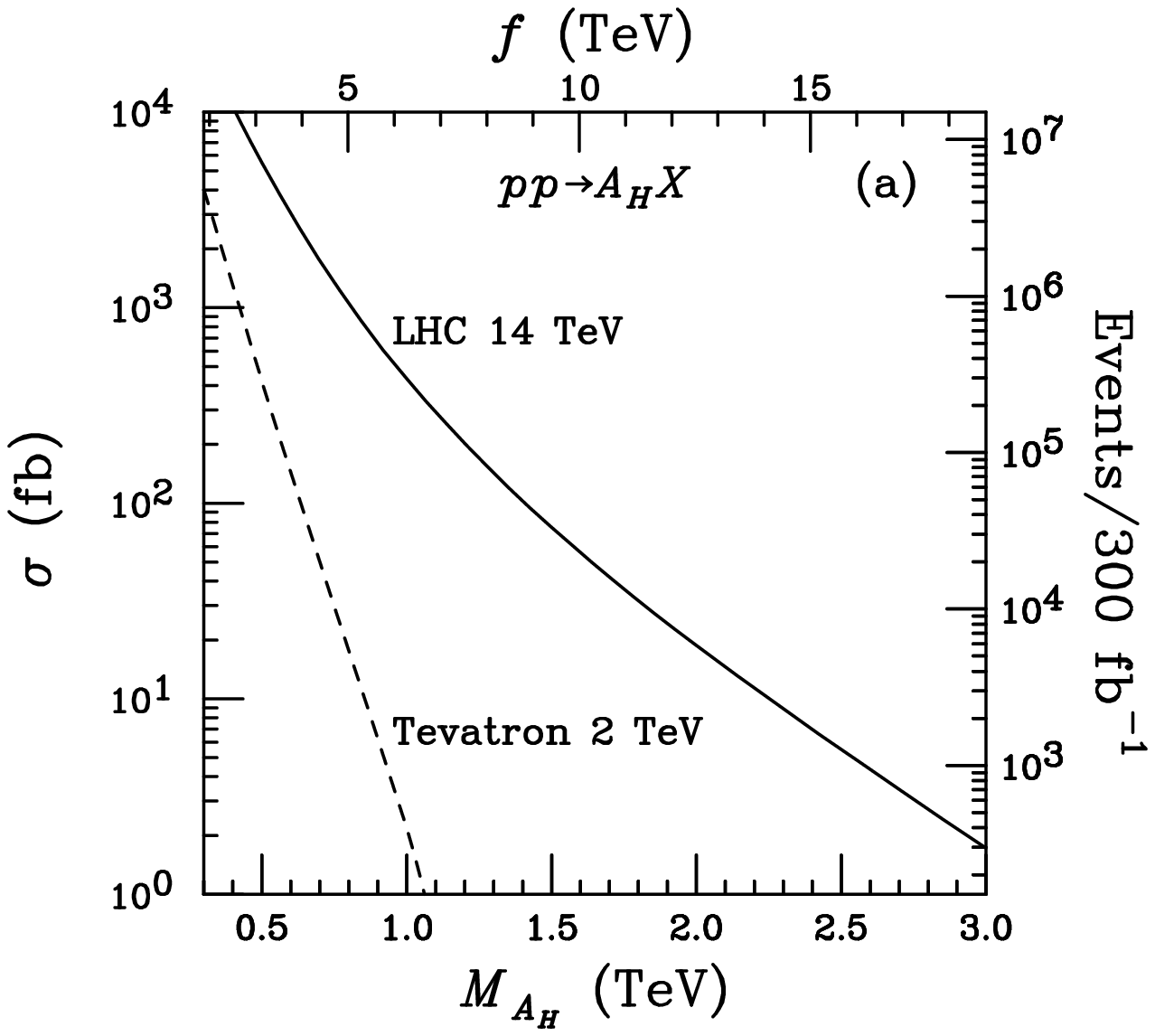}}
\end{center}
\caption{Cross section for $A_H$ production in Drell-Yan at the LHC
and Tevatron, for $\cot\theta^{\prime} = 1$.  From $^{11}$.}
\label{fig:sigmaAH}
\end{figure}

\begin{figure}
\begin{center}
\resizebox{10cm}{!}{\includegraphics{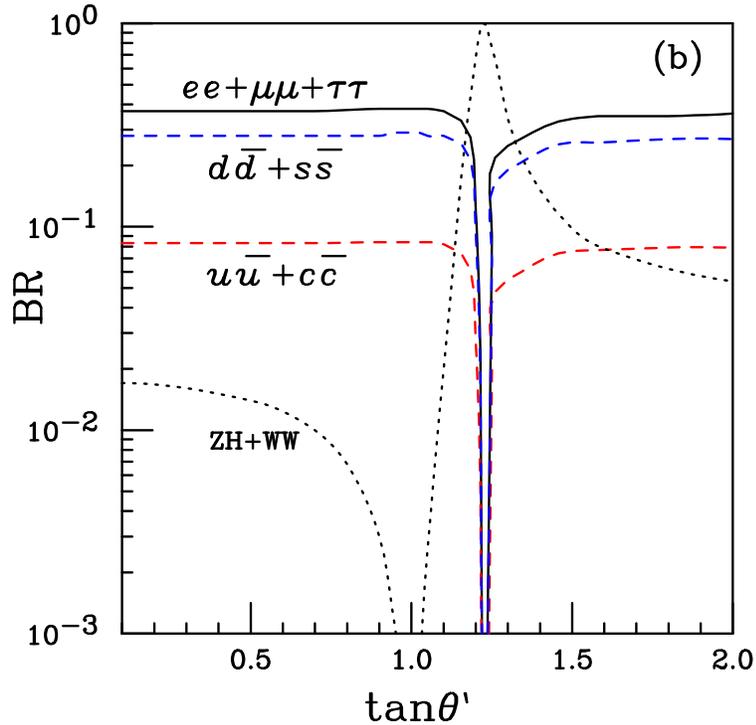}}
\end{center}
\caption{Branching ratios of $A_H$ into fermions and $ZH+WW$ as a function
of $\tan\theta^{\prime}$, neglecting final-state mass effects.}
\label{fig:brAH}
\end{figure}

{\it $A_H$:}
The heavy U(1) gauge boson is the lightest new particle in the Littlest
Higgs model.  Its couplings to fermions are more model dependent than
those of the heavy SU(2) gauge bosons, since they depend on the U(1) charges
of the fermions.  Even the presence of $A_H$ is model-dependent, since
one can remove this particle from the Littlest Higgs model by gauging only
one U(1) group (hypercharge) without adding a significant amount of 
fine-tuning.  Nevertheless, we show in Figs.~\ref{fig:sigmaAH} and
\ref{fig:brAH} the cross section and branching ratios\footnote{Here
again we correct an error in \cite{LHpheno} in which the $A_H$ decay to
$W^+W^-$ was overlooked.}
of $A_H$ for the
anomaly-free choice of U(1) charges discussed in Sec.~\ref{sec:fermions}.

\begin{figure}
\begin{center}
\resizebox{10cm}{!}{\includegraphics{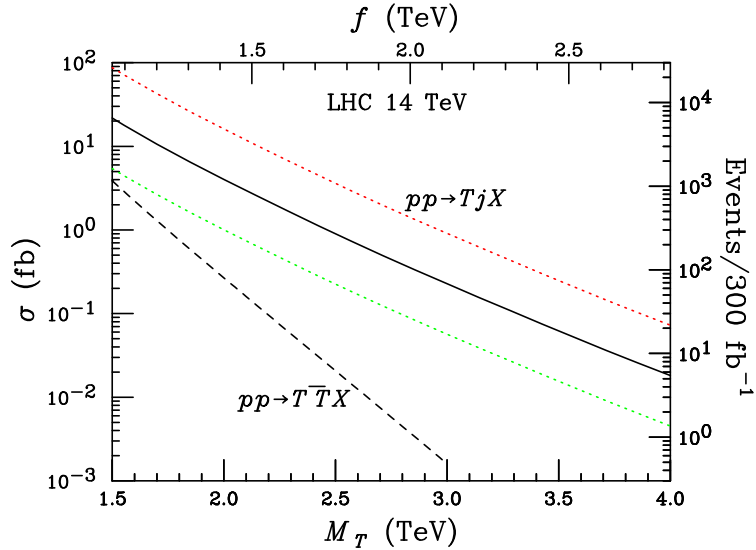}}
\end{center}
\caption{Cross sections for $T$ production at the LHC.  The single $T$
cross section is shown for $\lambda_1/\lambda_2 = 1$ (solid) and 
$\lambda_1/\lambda_2 = 2$ (upper dotted) and $1/2$ (lower dotted).
The QCD pair production cross section is shown for comparison (dashed).
From $^{11}$.}
\label{fig:toph}
\end{figure}

{\it $T$:}
The heavy top-partner $T$ can be pair produced by QCD interactions with
model-independent couplings.  However, this production mode is suppressed
by phase space due to the high mass of $T$.  The single $T$ production 
mode, $W^+b \to T$, is dominant for $M_T$ above about a TeV.  The 
cross section for single $T$ production depends on the ratio of couplings
$\lambda_1/\lambda_2$, which relates $M_T$ to the scale $f$.
The cross sections are shown in 
Fig.~\ref{fig:toph}.
The top-partner $T$ decays into $tH$, $tZ$ and $bW$ with partial widths 
in the ratio $1:1:2$.  

The top sector is quite similar in many of the other little Higgs models
in the literature, so that these general features of $T$ production and
decay should apply.  Some models contain more than one 
top-partner \cite{SU6Sp6,KaplanSchmaltz,custodialmoose,Spencer,Anntalk}, or
contain partners for the two light generations of fermions as well
\cite{KaplanSchmaltz,SkibaTerning}; in 
this case the phenomenology will be modified.

{\it $\Phi^{++}$:}
The doubly charged Higgs triplet state $\Phi^{++}$ can be singly 
produced through resonant $W^+W^+ \to \Phi^{++} \to W^+W^+$.  The 
cross section for this process is proportional to $v^{\prime 2}$, which
may make it difficult to see due to lack of rate.  The doubly
charged Higgs could also be pair produced if it is not too heavy.
The doubly charged Higgs can in principle decay to a pair of like-sign
charged leptons via the dimension-four operator $L\Phi L$, offering a 
more distinctive signature; however, the coupling is highly model 
dependent and care must be taken to avoid generating too large a neutrino
mass from the triplet vev.

\subsection{Future precision measurements}

As already explained, little Higgs models modify the precision electroweak
observables, so that a significant improvement in the measurements of 
these observables (at, for example, a ``Giga-$Z$'' machine) should 
turn up a signal.  However, there are other precision measurements 
in which the little Higgs should show its effects.  Here we 
discuss the effects of the Littlest Higgs model on triple gauge couplings
\cite{LHpheno} and loop-induced Higgs decays into gluon and photon pairs
\cite{LHloop}.  The effects of the Littlest Higgs model on
$b \to s \gamma$ \cite{bsgamma}, the muon anomalous magnetic moment 
\cite{gminus2}, and double-Higgs production via gluon fusion \cite{doubleH}
have also been studied in the literature.

{\it Triple gauge couplings:}
The $WWZ$ triple gauge boson coupling in the Littlest
Higgs model is modified from its SM form due to the modification of
$G_F$ by $W_H$ exchange:
\begin{equation}
g_1^Z = \kappa_Z =
        1 + \frac{1}{\cos 2 \theta_W} \left\{
        \frac{v^2}{8f^2} \left[ -4 c^2 s^2
        + 5 (c^{\prime 2}-s^{\prime 2})^2 \right]
        - \frac{2 v^{\prime 2}}{v^2} \right\},
\end{equation}
where the form-factors are defined according to
\begin{eqnarray}
\mathcal{L}_{WWV} &=& i g_{WWV} \left[
        g_1^V (W^+_{\mu\nu} W^{- \mu} - W^{+ \mu} W^-_{\mu\nu} ) V^{\nu}
        + \kappa_V W^+_{\mu} W^-_{\nu} V^{\mu\nu}
	\right. \nonumber \\ && \left. 
        + \frac{\lambda_V}{m_W^2} W_{\mu}^{+\nu} W_{\nu}^{-\rho} V_{\rho}^{\mu}
        \right].
\end{eqnarray}
At present, the constraints from the $WWZ$ coupling are weak compared
to those from electroweak precision measurements.
However, at a future linear collider, a precision of $10^{-3}-10^{-4}$ 
on $g_1^Z$ and $\kappa_Z$ should be reachable; this would be sensitive 
to $f\sim (15-50)v\sim 3.5-12$ TeV for generic parameter values.
Unfortunately for this measurement, in the region of parameter space in
which the electroweak precision bounds are loosened (small $c$ and $v^{\prime}$
and $c^{\prime} \simeq s^{\prime}$) the modification to $g_1^Z$ and $\kappa_Z$
is also suppressed.

\begin{figure}
\begin{center}
\resizebox{10cm}{!}{
	\rotatebox{270}{\includegraphics{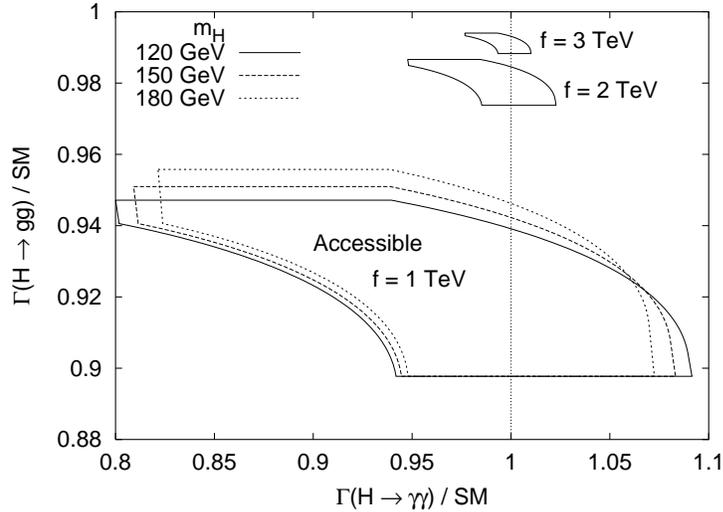}}}
\end{center}
\caption{Range of values of $\Gamma(H \to gg)$ versus 
$\Gamma(H \to \gamma\gamma)$ accessible in the Littlest Higgs model
normalized to the SM value, for $m_H = 120$, 150, 180 GeV and 
$f = 1$, 2, 3 TeV.  From $^{12}$.}
\label{fig:PWblob}
\end{figure}

{\it Loop-induced Higgs decays:}
The Higgs decays into gluon pairs or photon
pairs will be modified in the Littlest Higgs model by the new particles 
running in the loop and by the shifts in the Higgs couplings
to the SM $W$ boson and top quark due to the structure of the
model.  These modifications of the Higgs couplings to gluon or photon pairs
scale like $1/f^2$, and thus decouple at high $f$ scales.
The range of partial widths as a function of $f$ accessible
by varying the other model parameters are shown in Fig.~\ref{fig:PWblob}.

Are these corrections observable?  For $f \geq 1$ TeV, the correction
to $\Gamma(H \to gg)$ is always less than 10\%.  This is already smaller
than the remaining SM theoretical uncertainty on the gluon fusion cross 
section due to uncalculated higher-order QCD corrections.
For the partial width to photons, the situation is more promising because
the QCD corrections are well under control.  At the LHC, the 
$H \to \gamma\gamma$ decay rate can be measured to 15-20\%; this probes
$f < 1.0$ TeV at $1\sigma$.
A linear $e^+e^-$ collider has only comparable precision since the
$H \to \gamma\gamma$ branching ratio measurement is limited by statistics.
The most promising measurement would be done at a photon collider,
in which the $\gamma\gamma \to H \to b \bar b$ rate could be measured 
to about 2\%.  Combining this with a measurement of the branching ratio
of $H \to b \bar b$ to about 1.5-2\% at the $e^+e^-$ collider allows
the extraction of $\Gamma(H \to \gamma\gamma)$ with a precision of about 
3\%.  Such a measurement would be sensitive to $f < 2.7$ TeV at the $1\sigma$
level, or $f < 1.8$ TeV at the $2\sigma$ level.  A $5\sigma$ deviation is
possible for $f < 1.2$ TeV.  
For comparison, the electroweak precision constraints require
$f \gsim 1$ TeV in the Littlest Higgs model.

The biggest model dependence in the loop-induced Higgs decays in little
Higgs models comes from the content of the Higgs sector at the electroweak
scale.  In models with only one light Higgs doublet, our general conclusions
should hold, up to factors related to the multiplicity and detailed 
couplings of the new heavy particles.  However, many little Higgs models
\cite{minmoose,SU6Sp6,KaplanSchmaltz,custodialmoose,SkibaTerning}
contain two light Higgs doublets.  In this case, mixing between the 
two neutral CP-even Higgs particles and the contribution of a
relatively light charged Higgs boson running in the loop can lead to
large deviations in the couplings of the lightest Higgs boson to gluon
or photon pairs, swamping the effects from the heavy states.

\section{\label{sec:concl} Outlook and Conclusions}

The little Higgs idea provides a new way to address the little hierarchy
problem of the Standard Model by making the Higgs a pseudo-Nambu-Goldstone 
boson of a spontaneously broken global symmetry.  The global symmetry
is explicitly broken by gauge and Yukawa interactions; however, no 
single interaction breaks all the symmetry protecting the Higgs mass.
This prevents quadratically divergent radiative corrections to the Higgs
mass from appearing at the one-loop level, and thus allows the
cutoff scale to be pushed higher by one loop factor, to $\sim 10$ TeV.

From the bottom-up point of view, the quadratically divergent radiative 
corrections to the Higgs mass due to top quark, gauge boson, and Higgs 
loops are canceled by new heavy quarks, gauge bosons and scalars, 
respectively.  In contrast to supersymmetry, the cancellations occur
between loops of particles with the {\it same} statistics.  

The details of the phenomenology depend on the specific model.
Since quite a few little Higgs models have already appeared on the
market over the past two years, finding generic features of the 
phenomenology is important.
Very generically, there must be new gauge bosons, fermions and scalars
to cancel the quadratic divergences in the Higgs mass.
Less generically, models with product gauge groups
of the form [SU(2)$\times$U(1)]$^2$ contain an SU(2) triplet of new heavy
gauge bosons, $Z_H,W^{\pm}_H$.

There is some tension between the precision electroweak constraints
pushing up the new particle masses and the requirement that the new 
particles be light to avoid fine tuning. 
However, by tuning the parameters of the models appropriately one can
satisfy both constraints.  This tuning of the parameters should be
explained in the ultraviolet completion of the nonlinear sigma model.
Our developing understanding of the effects of
little Higgs models on the electroweak precision observables is now
driving model building to incorporate features that loosen the constraints.
Taking these constraints into account,
the new particles should live in the $1-2$ TeV mass range 
and should be accessible at the LHC.

\section*{Acknowledgments}
It is a pleasure to thank Tao Han, Bob McElrath and Lian-Tao Wang 
for collaborations leading to the papers \cite{LHpheno,LHloop} on
which this talk was based.
I also thank the organizers of the SUGRA-20 conference for inviting
me to participate.
My research is supported in part by the U.S.~Department of Energy
under grant DE-FG02-95ER40896
and in part by the Wisconsin Alumni Research Foundation.



\begin{thebibliography}{99}

\bibitem{bigmoose}
N.~Arkani-Hamed, A.~G.~Cohen and H.~Georgi,
Phys.\ Lett.\ B {\bf 513}, 232 (2001)
[arXiv:hep-ph/0105239].

\bibitem{Jaytalk}
J.~G.~Wacker,
arXiv:hep-ph/0208235.

\bibitem{PNGBhiggs}
S.~Dimopoulos and J.~Preskill,
Nucl.\ Phys.\ B {\bf 199}, 206 (1982);
D.~B.~Kaplan and H.~Georgi,
Phys.\ Lett.\ B {\bf 136}, 183 (1984);
D.~B.~Kaplan, H.~Georgi and S.~Dimopoulos,
Phys.\ Lett.\ B {\bf 136}, 187 (1984);
H.~Georgi and D.~B.~Kaplan,
Phys.\ Lett.\ B {\bf 145}, 216 (1984);
H.~Georgi, D.~B.~Kaplan and P.~Galison,
Phys.\ Lett.\ B {\bf 143}, 152 (1984);
M.~J.~Dugan, H.~Georgi and D.~B.~Kaplan,
Nucl.\ Phys.\ B {\bf 254}, 299 (1985);
T.~Banks,
Nucl.\ Phys.\ B {\bf 243}, 125 (1984).

\bibitem{minmoose}
N.~Arkani-Hamed, A.~G.~Cohen, E.~Katz, A.~E.~Nelson, 
T.~Gregoire and J.~G.~Wacker,
JHEP {\bf 0208}, 021 (2002)
[arXiv:hep-ph/0206020].

\bibitem{Littlest}
N.~Arkani-Hamed, A.~G.~Cohen, E.~Katz and A.~E.~Nelson,
JHEP {\bf 0207}, 034 (2002)
[arXiv:hep-ph/0206021].

\bibitem{SU6Sp6}
I.~Low, W.~Skiba and D.~Smith,
Phys.\ Rev.\ D {\bf 66}, 072001 (2002)
[arXiv:hep-ph/0207243].

\bibitem{KaplanSchmaltz}
D.~E.~Kaplan and M.~Schmaltz,
arXiv:hep-ph/0302049.

\bibitem{custodialmoose}
S.~Chang and J.~G.~Wacker,
arXiv:hep-ph/0303001.

\bibitem{SkibaTerning}
W.~Skiba and J.~Terning,
arXiv:hep-ph/0305302.

\bibitem{Spencer}
S.~Chang,
arXiv:hep-ph/0306034.

\bibitem{LHpheno}
T.~Han, H.~E.~Logan, B.~McElrath and L.~T.~Wang,
Phys.\ Rev.\ D {\bf 67}, 095004 (2003)
[arXiv:hep-ph/0301040].

\bibitem{LHloop}
T.~Han, H.~E.~Logan, B.~McElrath and L.~T.~Wang,
Phys.\ Lett.\ B {\bf 563}, 191 (2003)
[arXiv:hep-ph/0302188].

\bibitem{Anntalk}
A.~E.~Nelson,
arXiv:hep-ph/0304036.

\bibitem{GrahamEW1}
C.~Csaki, J.~Hubisz, G.~D.~Kribs, P.~Meade and J.~Terning,
Phys.\ Rev.\ D {\bf 67}, 115002 (2003)
[arXiv:hep-ph/0211124].

\bibitem{JoAnneEW}
J.~L.~Hewett, F.~J.~Petriello and T.~G.~Rizzo,
arXiv:hep-ph/0211218.

\bibitem{GrahamEW2}
C.~Csaki, J.~Hubisz, G.~D.~Kribs, P.~Meade and J.~Terning,
arXiv:hep-ph/0303236.

\bibitem{JayEW}
T.~Gregoire, D.~R.~Smith and J.~G.~Wacker,
arXiv:hep-ph/0305275.

\bibitem{gustavo}
G.~Burdman, M.~Perelstein and A.~Pierce,
Phys.\ Rev.\ Lett.\  {\bf 90}, 241802 (2003)
[arXiv:hep-ph/0212228].

\bibitem{bsgamma}
W.~j.~Huo and S.~h.~Zhu,
arXiv:hep-ph/0306029.

\bibitem{gminus2}
S.~C.~Park and J.~h.~Song,
arXiv:hep-ph/0306112.

\bibitem{doubleH}
C.~Dib, R.~Rosenfeld and A.~Zerwekh,
arXiv:hep-ph/0302068.

\end{thebibliography}
\end{document}